\documentclass{article}
\usepackage{spconf,amsmath,graphicx}
\usepackage{graphicx,amsfonts,amsmath,amssymb,xcolor,ntheorem}
\usepackage{algorithm,algorithmic}
\usepackage{multirow}
\usepackage[noadjust]{cite}
\def\comment#1{}

\newcommand{\Dmat}{{\bf D}}

\newcommand{\Imat}{{\bf I}}

\newcommand{\Rmat}[0]{{{\bf R}}}

\newcommand{\Tmat}[0]{{{\bf T}}}

\newcommand{\Xmat}{{\bf X}}
\newcommand{\Ymat}[0]{{{\bf Y}}}

\newcommand{\av}{\boldsymbol{a}}

\newcommand{\wv}{\boldsymbol{w}}

\newcommand{\xv}{\boldsymbol{x}}
\newcommand{\yv}{\boldsymbol{y}}

\newcommand{\zv}{\boldsymbol{z}}

\newcommand{\Phimat}{\boldsymbol{\Phi}}

\newcommand{\thetav}{\boldsymbol{\theta}}

\newcommand{\ts}{^{\top}}

\newcommand{\ie}{{\em i.e.}}
\newcommand{\TV}{{\rm TV}}

\title{Generalized Alternating Projection Based Total Variation Minimization for Compressive Sensing}
\name{Xin Yuan}
\address{Bell Labs, Alcatel-Lucent, 600 Mountain Avenue, Murray Hill, NJ, 07974, USA}
\begin{document}
%
\maketitle
\begin{abstract}
We consider the total variation (TV) minimization problem used for compressive sensing and solve it using the generalized alternating projection (GAP) algorithm.
Extensive results demonstrate the high performance of proposed algorithm on compressive sensing, including two dimensional images, hyperspectral images and videos.
We further derive the Alternating Direction Method of Multipliers (ADMM) framework with TV minimization for video and hyperspectral image compressive sensing under the CACTI and CASSI framework, respectively. 
Connections between GAP and ADMM are also provided. 
\end{abstract}
\section{Introduction}
The generalized alternating projection (GAP) algorithm, originally proposed in~\cite{Liao14GAP}, has demonstrated excellent performance on diversely compressive sensing (CS) problems, including two dimensional (2D) images~\cite{Liao14GAP,Yuan15Lensless,Yuan14TSP,Yuan15GMM,Yuan15GAP}, hyperspectral images~\cite{Yuan15JSTSP,Tsai15OL}, videos~\cite{Llull13COSI,Patrick13OE, Yang14GMM,Yang14GMMonline,Tsai15COSI,Yuan15FiO,Stevens15ASCI,CS15Book}, depth images~\cite{Yuan14CVPR,Llull14COSI,Llull15Optica}, and polarization images~\cite{Tsai15OE}. However, all the above demonstrations, including the real system applications, utilize GAP to solve the compressive sensing problem in the transformation domain, \ie, the wavelet or DCT (Discrete Cosine Transformation) domain. Specifically, GAP is used to solved the following problem:
\begin{equation}
\min_{\wv} \|\wv\|_{\ell_{2,1}^{\cal G \beta}}, ~~ \text{subject to}~~ \Phimat \xv = \yv, ~~ (\text{with}~~ \xv = \Tmat \wv),
\end{equation}
where $\Tmat$ is the transformation matrix, which is a Kronecker product of different bases if the transformation is performed on each dimension. $\yv$ is the measurement, and $\Phimat$ is the sensing matrix. $\xv$ is the desired signal and $\wv$ denotes the corresponding coefficients in the transformed domain.  $\|\cdot\|_{\ell_{2,1}^{\cal G \beta}}$ signifies the weighted group $\ell_{2,1}$ norm~\cite{Liao14GAP} defined as
$\|\wv\|_{\ell_{2,1}^{\cal G \beta}} = \sum_{k=1}^K \beta_k\|\wv_{{\cal G}_k}\|_2$,
with $\|\cdot\|_2$ denoting the $\ell_2$ norm and $\wv_{{\cal G}_k}$ a sub-vector of $\wv$ containing components indexed by ${\cal G}_k$ with $\beta_k$ denoting the weight for this group.
Though great success has been obtained, it is not always easy to find a good transformation $\Tmat$ to be imposed on the data. Furthermore, this transformation will take some computational time and the selection of groups and weights affects the results significantly~\cite{Yuan14CVPR}.

On the other hand, the total variation (TV) based algorithms have demonstrated high performance on variously CS problems~\cite{Bioucas-Dias2007TwIST,Wang08TV,Yang10TV,Li13COA,Huang13ICIP,Jiang14APSIPA}. It solves the following problem:
\begin{equation} \label{eq:TV}
\min_{\xv} \|\TV(\xv)\|, ~~ \text{subject to}~~ \Phimat \xv = \yv,
\end{equation}
where $\|\TV(\xv)\|$ denotes the TV norm.
The question is can we solve the TV minimization problem in (\ref{eq:TV}) by GAP? This paper will fill the research gap by developing a new algorithm using GAP to solve (\ref{eq:TV}).

\section{Mathematic Formulation}
In the following, we use $\|\TV(\xv)\| = \|\Dmat \xv\|_1$, where $\Dmat$ is the differential operation and it will be performed on different dimensions if $\xv$ denotes an image or video.

Under the GAP formulation, the problem in \eqref{eq:TV} can be rewritten as:
\begin{eqnarray} \label{eq:TVball}
\min_{\xv, C} ~~\text{subject to}~~ \|\TV(\xv)\|\le C ~~ \text{and}~~ \Phimat \xv = \yv.
\end{eqnarray}
where $C$ is the radius of the $\ell_1$-ball based on the TV of the signal.

We solve \eqref{eq:TVball} as a series of alternating projection problem:
\begin{eqnarray}
\left(\xv^{(t)}, \thetav^{(t)}\right) &=& \arg\min_{\xv,\thetav} \frac{1}{2}\|\xv-\thetav\|_2,\nonumber\\
\text{subject to}~~ \|\TV(\thetav)\|&\le& C^{(t)}~~\text{and}~~ \Phimat\xv = \yv,
\end{eqnarray}
which is equivalent to
\begin{eqnarray} \label{eq:TVregu}
\left(\xv^{(t)}, \thetav^{(t)}\right) &=&\arg \min_{\xv,\thetav} \frac{1}{2}\|\xv-\thetav\|_2^2 + \lambda \|\TV (\thetav)\|, \nonumber \\
\text{subject to}&& \Phimat\xv = \yv,
\end{eqnarray}
where $\lambda$ is a regulizer and $t$ denotes the iteration number.

\section{GAP-TV algorithm}
Equation~\eqref{eq:TVregu} can be solved by alternating updating $\thetav$ and $\xv$. Given $\thetav$, the update of $\xv$ is simply an Euclidean projection of $\thetav$ on the linear manifold; this is solved by
\begin{eqnarray}\label{eq:gap_xt}
\xv^{(t)}&=& \thetav^{(t-1)} + \Phimat\ts(\Phimat\Phimat\ts)^{-1}(\yv - \Phimat\thetav^{(t-1)}), \label{eq:xvt}
\end{eqnarray}
where we assume $\Phimat\Phimat\ts$ is invertible and in a lot of CS applications~\cite{Huang13ICIP,Yuan14CVPR,Yuan15JSTSP,Yuan15FiO,Yuan14Tree}, $\Phimat\Phimat\ts$ is a diagonal matrix and it is very easy to calculate the inversion.

Given $\thetav$, the update of $\xv$ is an TV denoising problem which can be solved by the iterative clipping algorithm~~\cite{Beck09TV,zhu08Dual}:
\begin{eqnarray}
\thetav^{(t)}&=& \xv^{(t)} - \Dmat\ts \zv^{(t)}, \label{eq:thetavt}\\
\zv^{(t)} &=& {\rm clip}\left(\zv^{(t-1)} + \frac{1}{\alpha} \Dmat \thetav^{(t-1)}, \frac{\lambda}{2}\right), \label{eq:zvt}
\end{eqnarray}
where $\zv^{(0)} = 0$ and $\alpha\ge {\rm maxeig}(\Dmat\Dmat\ts)$ and the clipping function ${\rm clip} (\cdot)$ is defined as:
\begin{equation}
{\rm clip}(b, T):=\left\{\begin{array}{ll}
b, & {\text {if}  }~~|b|\le T,\\ T~ {\rm sign}(b), & {\text {otherwise}  }.
\end{array}\right.
\end{equation}
The iteration starts from $\zv^{(0)} = 0$, then $\thetav$, and finally update $\xv$. 

	\subsection{The Accelerated Update for $\xv$}
	According to~\cite{Liao14GAP}, the linear manifold can also be adaptively adjusted and so \eqref{eq:xvt} can be modified as
	\begin{eqnarray}
	\xv^{(t)}&=& \thetav^{(t-1)} + \Phimat\ts(\Phimat\Phimat\ts)^{-1}(\yv^{(t-1)} - \Phimat\thetav^{(t-1)}),
	\end{eqnarray}
	with
	\begin{eqnarray}
	\yv^{(t)}&=& \yv^{(t-1)} + (\yv - \Phimat \thetav^{(t-1)}), \forall t\ge 1,\\
	\text{or}~~ \yv^{(t)}&=& \yv+ \Delta(\yv - \Phimat \thetav^{(t-1)}), \forall t\ge 1, \Delta\ge 0.
	\end{eqnarray}
	
	\subsection{Relation to ADMM}
	\label{Sec:ADMM_TV}
	The Alternating Direction Method of Multipliers (ADMM) algorithm~\cite{ADMM2011Boyd} provides an alternative solution to a lot of optimization problems.
	Though both ADMM and GAP introduce an auxiliary variable, \ie, $\thetav$ in (\ref{eq:TVregu}) to solve the problem, the difference lies in how to update $\xv$.
	Under the ADMM formulation, introducing another regulizer $\eta$, the cost function will be:
	\begin{align}
	L(\thetav,\xv, \lambda,\eta) &= \frac{1}{2}\|\yv-\Phimat\xv\|_2^2 + \frac{\eta}{2}\|\xv-\thetav\|_2^2 + \lambda \|\TV (\thetav)\|.
	\end{align}
	ADMM cyclically solves the following subproblems:
	\begin{eqnarray}
	\xv^{(t+1)}&: =& \arg \min_{\xv} \frac{1}{2}\|\yv-\Phimat\xv\|_2^2 + \frac{\eta}{2}\|\xv-\thetav^{(t)}\|_2^2,  \label{eq:admm_x_t+1}\\ 
	\thetav^{(t+1)} &:= & \arg\min_{\thetav} \frac{\eta}{2}\|\xv^{(t+1)}-\thetav\|_2^2 + \lambda \|\TV (\thetav)\|. \label{eq:admm_theta_t+1}
	\end{eqnarray}
	Given $\thetav$, (\ref{eq:admm_x_t+1}) is a quadratic optimization problem of $\xv$ and can be simplified to:
	\begin{eqnarray}
	(\Phimat\ts\Phimat + \eta\Imat)\xv = \Phimat\ts \yv + \eta \thetav,
	\end{eqnarray}
	which admits the following closed-form solution:
	\begin{eqnarray}
	\xv^{(t+1)}&=& (\Phimat\ts\Phimat + \eta\Imat)^{-1} (\Phimat\ts \yv + \eta \thetav^{(t)}).
	\end{eqnarray}
	Since in the CS framework, $\Phimat$ is a fat matrix, $(\Phimat\ts\Phimat + \eta\Imat)$ will be a large matrix and thus the matrix inversion formula can be used to simplify the problem:
	\begin{align} \label{eq:admm_xv_inv}
	\xv^{(t+1)} &= \left[\eta^{-1}{\Imat}-\eta^{-1}\Phimat^{\top}(\Imat + \Phimat\eta^{-1}\Phimat^{\top})^{-1}\Phimat\eta^{-1}\right]\nonumber\\
	&\times[\Phimat^{\top}\yv + \eta \thetav^{(t)}].
	\end{align}
	With some special constraints on the sensing matrix $\Phimat$, (\ref{eq:admm_xv_inv}) can be simplified as follows:
	
		a) For the 2D image CS considered in this paper, the sensing matrix is the permuted Hadamard matrix, which has been implemented in the lensless camera~\cite{Huang13ICIP}.
		In this case $\Phimat\Phimat\ts = \Imat$, therefore,
		\begin{eqnarray}
		\xv^{(t+1)} &= & \thetav^{(t)} + \frac{\Phimat\ts (\yv- \Phimat \thetav^{(t)})}{\eta + 1},
		\end{eqnarray}
		which is same as (\ref{eq:gap_xt}) if $\eta = 0$.
		However, in the experiments, we observed that tuning this $\eta$ is critical to the performance of the algorithm.
		
		b) For the video~\cite{Patrick13OE,Yuan14CVPR} and hyperspectral image~\cite{Yuan15JSTSP} CS considered in this paper, the sensing matrix is a shifting binary mask and $\Phimat\Phimat\ts$ is a diagonal matrix:
		\begin{eqnarray}
		\Phimat\Phimat\ts &=& {\rm diag}\{r_1, \dots, r_M\} \stackrel{\rm def}{=} \Rmat,
		\end{eqnarray}
		where $M$ is the number of rows in $\Phimat$ and for the video and hyperspectral image CS, it is the pixel number in each frame (or the measurement).
		Following this, in (\ref{eq:admm_xv_inv}):
		\begin{eqnarray}
		(\Imat + \Phimat\eta^{-1}\Phimat^{\top})^{-1}&\stackrel{\rm def}{=}& \tilde{\Rmat}
		\end{eqnarray}
		is also a diagonal matrix with diagonal elements:
		\begin{equation}
		\tilde{\Rmat} = {\rm diag}\left\{\frac{\eta }{\eta+r_1}, \dots, \frac{\eta }{\eta+r_M} \right\},
		\end{equation}
		and
		\begin{equation}
		\tilde{\Rmat} \Rmat = {\rm diag}\left\{\frac{\eta~r_1}{\eta+r_1}, \dots, \frac{\eta ~r_M}{\eta+r_M} \right\}.
		\end{equation}
		
		In this case, $\xv^{(t+1)}$ can be update by:
		\begin{eqnarray}
		\xv^{(t+1)} &= & \thetav^{(t)} + \frac{\Phimat\ts(\eta\Imat - \tilde{\Rmat}\Rmat)\yv}{\eta^2} - \frac{\Phimat\ts \tilde{\Rmat} \Phimat \thetav}{\eta}
		\end{eqnarray}
		with 
		\begin{equation}
		(\eta\Imat - \tilde{\Rmat}\Rmat) = {\rm diag}\left\{\frac{\eta^2}{\eta+r_1}, \dots, \frac{\eta^2}{\eta+r_M} \right\}.
		\end{equation}
		Following this,
		\begin{eqnarray}
		\xv^{(t+1)} &= & \thetav^{(t)} + \Phimat\ts \av,
		\end{eqnarray}
		where $\av$ can be calculated element-wise via:
		\begin{eqnarray}
		a_m &=& \frac{[\yv-\Phimat\thetav^{(t)}]_m}{\eta + r_m},
		\end{eqnarray}
		with $[\cdot]_m$ denoting the $m$-th element of the vector inside $[~]$.
		Again, the update for $\xv$ will be the same as GAP if $\eta=0$.
		
		Therefore, $\xv^{(t+1)}$ can be calculated very efficiently.
		Given $\xv$, (\ref{eq:admm_theta_t+1}) is a TV denoising problem and can be solved by the iterative clipping algorithm in (\ref{eq:thetavt})-(\ref{eq:zvt}).
		
		In summary, the difference of GAP and ADMM lies in the updating equation of $\xv$, where the Euclidean projection is used in GAP while ADMM introduces a new regulizer $\eta$. We have observed this $\eta$ affects the results critically in the experiments.
		One important property of GAP compared with other algorithms, {\em e.g.}, ADMM, TwIST~\cite{Bioucas-Dias2007TwIST}, TVAL3~\cite{Li13COA}, {\em et. al.},  is that no parameter is required to tune to update this $\xv$, since it is only an Euclidean projection.

\section{Simulation}
\subsection{CS of 2D image}
We test the compressive sensing inversion using GAP-TV on diverse 2D images used in~\cite{Yuan15GMM,Dong14TIP}, with the compressive sensing matrix defined by the permuted Hadamard matrix as used in the real lensless camera~\cite{Huang13ICIP}. All the images are resized to $256\times 256$ and the CSr is defined as:
\begin{equation}
\text{CSr} = \frac{{\text {number of rows in }} \Phimat}{{\text {number of columns in }} \Phimat},
\end{equation}
where the number of columns in $\Phimat$ is equivalent to the entire pixel number of the image.
We not only test the high CSr case, \ie, CSr $\in [0.1~ 0.5]$, but also test the extremely low CSr case, \ie, CSr $\in [0.02~0.1]$, which may be of interest to the real applications that are used to detect unusual events~\cite{Jiang12Inverse} without caring too much of the image quality. Results are summarized in Table~\ref{Table:sim_PSNR}, where `Acc-GAP-TV' denotes the accelerated GAP-TV and GAP-w signifies the GAP with wavelet transformation~\cite{Liao14GAP}.
\begin{table*}[htbp!]
	\caption{Reconstruction PSNR (dB) of different images with diverse algorithms at various CSr.}
	\centering{
		\small{
		\begin{tabular}{c|c|ccccccccccc}
			\hline Image & CSr
			& 0.02  & 0.04 & 0.05 & 0.06 & 0.07 & 0.08  & 0.1 & 0.2 & 0.3 & 0.4 & 0.5 \\
			\hline \hline
			& Acc-GAP-TV & \textbf{19.88} &\textbf{21.78} & \textbf{22.46} & \textbf{23.10} & \textbf{23.60} & \textbf{24.11} &  \textbf{24.82} & \textbf{27.37}& \textbf{29.10} & \textbf{30.60} & \textbf{32.01}\\
			& GAP-TV & 19.19  &	20.99&	21.65 &	22.28&	22.83 &	23.35  &	24.10& 	26.68&	28.36 &	29.69 &	30.93\\
			& ADMM-TV & 18.63 &	20.32 &	20.93&	21.51 &	22.03 &	22.54&		23.27&	25.74 &	27.34 &	28.51 &	29.58\\
			barbara	& GAP-w & 18.91 &	19.59&	21.25&	21.70&	21.61&	21.81 &	23.30&	25.49& 26.99& 	28.99&	30.51\\
			& TVAL3 & 18.92 &		20.45&	21.09 &	21.67&	22.21 &	22.75&		23.59&	26.39&	28.31 &	28.05 &	29.32\\
			& TwIST & 18.06 &		19.72&	20.06&	20.65 &	19.90&	20.79 &	21.93&	23.04&	25.30&	26.16&	26.93\\
			\hline
			& Acc-GAP-TV & \textbf{20.56} &	\textbf{22.49}&	\textbf{23.11}&	\textbf{23.64} &	\textbf{24.23}&	\textbf{24.70} &	\textbf{25.62}&	\textbf{29.08} &	\textbf{31.68} &	\textbf{33.84} &	\textbf{35.90}\\
			& GAP-TV & 19.89  &	21.73& 22.30&	22.79&	23.34&	23.78 &	24.60&	27.64 &	29.85 &	31.61 &	33.23\\
			& ADMM-TV & 19.31  &	21.09&	21.61&	22.08 &	22.54&	22.94 &	23.69 &	26.39&	28.29 &	29.82&	31.19\\
			boats & GAP-w & 19.59&		21.28 &	21.87&	22.35 &	22.87&	23.07 &	24.14 &	25.18 &	26.32& 25.25 &	27.30\\
			& TVAL3 & 19.45 &		20.89&	21.39&	21.84 &	22.34&	22.76&	23.59&	26.80 &	29.05&	28.92&	30.40\\
			& TwIST & 18.85 &		20.40 &	20.38 &	21.10 &	20.16 &	20.91&	22.21&	23.47&	25.77 &	26.67&	27.53\\
			\hline
			& Acc-GAP-TV & \textbf{19.32} & 		\textbf{21.10}&	\textbf{21.81} &	\textbf{22.36} &	\textbf{22.91 }&	\textbf{23.32}&	\textbf{24.26} &	\textbf{27.72} &	\textbf{30.25}&	\textbf{32.19}&	\textbf{33.69}\\
			& GAP-TV & 18.75  &	20.51&	21.20 &	21.72&	22.25&	22.64&	23.40&	26.57&	28.83&	30.49 &	31.87\\
			& ADMM-TV & 18.17 &	19.90&	20.53&	21.03 &	21.53 &	21.90&	22.57&	25.33&	27.39 &	28.91&	30.16\\
			cameraman & GAP-w & 18.67  &	18.97&	20.25 &	20.73 & 20.42 &	19.97 &		20.02&	21.42&	23.32&	24.67 &	28.66\\
			& TVAL3 & 18.64 &	19.91&	20.53 &	21.08&	20.83 &	22.07&	22.84&	26.33 &	29.07 & 	25.92 &	27.21\\
			& TwIST & 17.10 &	19.00&	18.97&	19.72 &	18.57 &	19.34&	20.44 &	21.44&	24.05 &	25.15 &	26.12\\
			\hline
			& Acc-GAP-TV & \textbf{24.54} &		\textbf{27.25}&	\textbf{28.18}&	\textbf{29.05} &	\textbf{29.78}&	\textbf{30.44}&		\textbf{31.68}&	\textbf{35.42}&	\textbf{38.04}&	\textbf{40.06} &\textbf{41.76}\\
			& GAP-TV & 23.31  &	25.95&	26.63&	27.32 &	27.89 &	28.47 &	29.02 &	33.03&	35.27&	36.89&	38.26\\
			& ADMM-TV & 22.23&	25.13&	25.76&	26.37 &	26.82&	27.29 &	28.20&	31.44&	33.50&	34.96&	36.16\\
			foreman & GAP-w & 21.22&	19.31&	23.84&	24.09&	26.08&	27.61 &		27.87 &	31.10&	33.39&	35.78&	36.59\\
			& TVAL3 & 22.55&	24.38&	24.95&	25.57&	26.12&	26.61&	27.63&	31.16&	33.79&	32.53&	33.89\\
			& TwIST & 20.41 &	23.70&	23.28&	24.69&	21.71 &	25.86&	26.13&	28.68&	30.64&	31.70&	32.56\\
			\hline
			& Acc-GAP-TV & \textbf{22.37}&		\textbf{24.97} &	\textbf{25.93}&	\textbf{26.89}&\textbf{	27.58}&	\textbf{28.27}&		\textbf{29.47}&	\textbf{33.05}&	\textbf{35.28}&	\textbf{37.06}&	\textbf{38.66}\\
			& GAP-TV & 21.55& 23.57&	24.40&	25.21&	25.84&	26.45&	27.55&	31.10&	33.22 &	34.77&	36.04\\
			& ADMM-TV & 20.97 &	22.74&	23.44&	24.10&	24.66&	25.18&	26.15&	29.53&	31.57&	32.98&	34.15\\
			house & GAP-w & 20.78&	18.85&	20.19&	20.99&	22.27&	21.37&	23.29&	27.51&	32.07&	33.36&	32.29\\
			& TVAL3 & 20.89&	22.38&	23.01&	23.57&	24.13&	24.65&	25.69&	29.52&	29.47&	31.01&	32.50\\
			& TwIST & 19.20&	20.85&	21.92&	22.69&	21.82&	22.69&	24.23&	25.60&	28.39&	29.48& 30.43\\
			\hline
			& Acc-GAP-TV & \textbf{20.77}&	\textbf{22.66}&	\textbf{23.40}&	\textbf{24.13}&	\textbf{24.75}&	\textbf{25.26}&		\textbf{26.15}&\textbf{	29.52}&	\textbf{32.17}&	\textbf{34.40}&	\textbf{36.23}\\
			& GAP-TV & 20.06&	21.75&	22.44& 23.13&	23.75&	24.24&	25.06&	28.08&	30.22&	31.98&	33.58\\
			& ADMM-TV & 19.51&	21.07&	21.67&	22.28&	22.86&	23.31&		24.05&	26.86&	28.70&	30.19&	31.54\\
			lena & GAP-w & 19.48&	19.30&	19.67&	20.53&	20.96&	20.62&		19.95&	24.92&	28.56&	30.42&	31.71\\
			& TVAL3 & 19.71&		21.09&	21.72&	22.36&	22.97&	23.47&		23.56&	27.69&	27.54&	28.99& 30.41\\
			& TwIST & 18.90&  	20.35& 	20.56&	21.34&	20.65 &	21.38&	22.44&	23.90&	26.26&	27.25&	28.08\\
			\hline
			& Acc-GAP-TV & \textbf{17.01}&	 \textbf{19.04}&	\textbf{20.17}&	\textbf{21.02}&	\textbf{21.83}&	\textbf{22.55}&		\textbf{23.85}&	\textbf{28.11}&	\textbf{31.24}&	\textbf{33.65}&	\textbf{35.30}\\
			& GAP-TV & 16.57& 	18.25&	19.19&	19.94&	20.69&	21.36&	22.57&	26.58&	29.26&	31.39&	33.01\\
			& ADMM-TV & 16.22&		17.59&	18.36&	19.00&	19.66 &	20.26&	21.40&	25.17&	27.66&	29.55&	31.03\\
			monarch & GAP-w & 15.74&	16.90&	17.70&	18.44&	18.97&	18.28&		18.84&	21.80&	23.63&	26.79&	28.38\\
			& TVAL3 & 16.40&		17.76&	18.51&	19.20&	19.96&	20.68&		22.12&	26.82&	29.75&	32.12&	27.93\\
			& TwIST & 15.68&	16.90&	17.56&	18.14&	18.55&	18.44&		19.52&	21.19&	24.36&	26.11&	27.32\\		
			\hline
			& Acc-GAP-TV & \textbf{20.85}&		\textbf{23.01}&	\textbf{23.69}& \textbf{24.32}& 	\textbf{25.03}&	\textbf{25.70}&		\textbf{26.96}&	\textbf{31.10}&	\textbf{33.80}&	\textbf{35.78}&	\textbf{37.12}\\
			& GAP-TV & 20.17&	22.25&	22.84&	23.30&	23.84&	24.34&	25.39&	29.11&	31.62&	33.42&	34.79\\
			& ADMM-TV & 19.49 &	21.60&	22.16&	22.59&	23.03&	23.44&	24.26&	27.47&	29.80&	31.48&	32.82\\
			parrot & GAP-w & 15.90& 	17.50&	18.93&	18.95&	18.90&	19.47&		19.66&	22.17&	24.78&	25.57&	25.30\\
			& TVAL3 & 19.94&	21.41&	21.92&	22.35 &	22.86&	23.30&		24.21&	28.55&	31.69&	33.93&	35.78\\
			& TwIST & 18.32&	20.66& 20.48&	21.27&	19.59&	20.54&		22.13&	23.01&	26.14&	27.44&	28.53\\		
			\hline \hline
		\end{tabular}}}
		\label{Table:sim_PSNR}
	\end{table*}
\vspace{-2mm}
\subsection{Video CS}
\label{Sec:video_cs}
\begin{figure}[htbp]
	\centering
	\vspace{-3mm}
	\includegraphics[width=.4\textwidth]{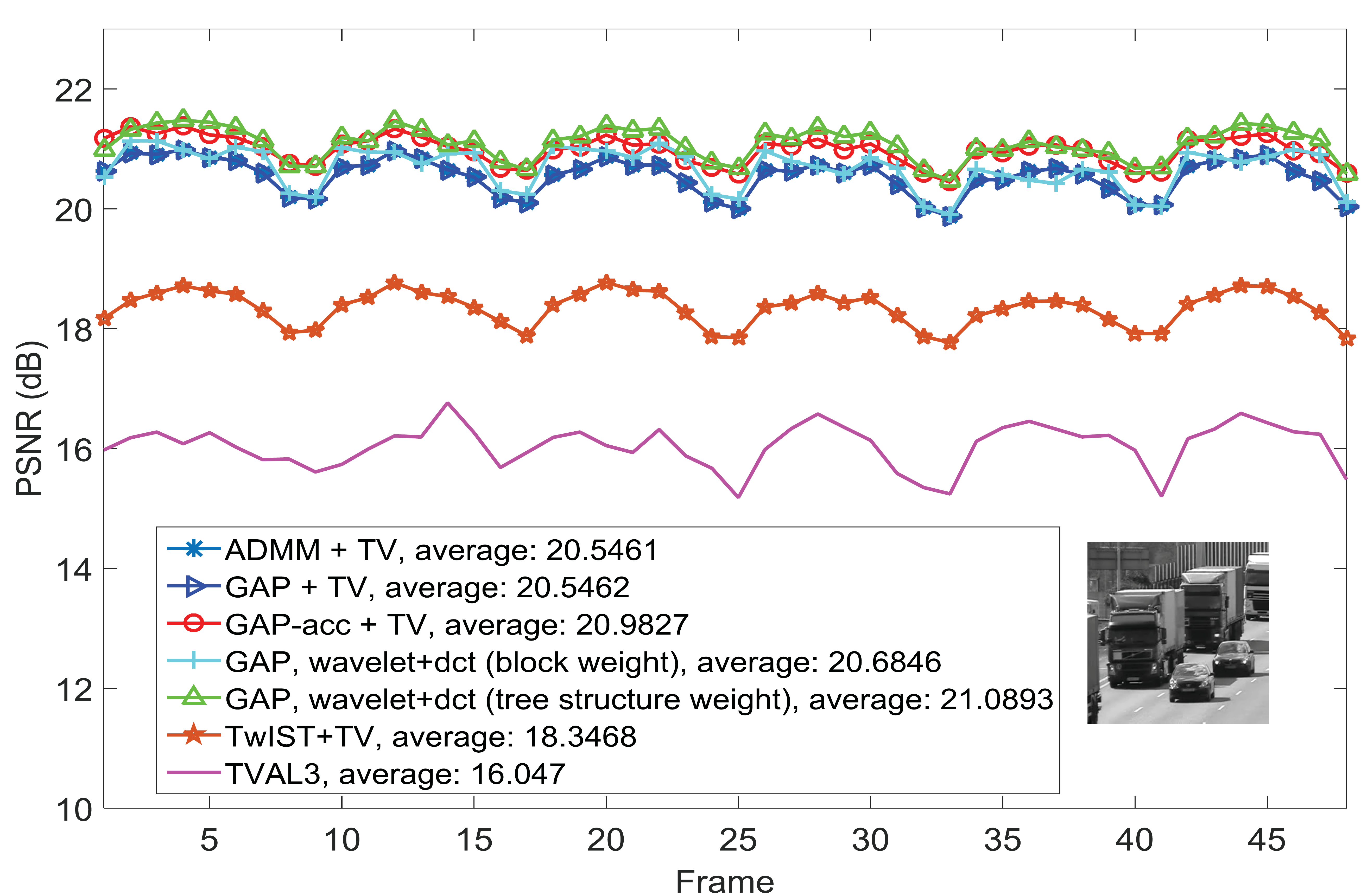}
	\vspace{-3mm}
	\caption{\small {Video compressive sensing (traffic) with different reconstruction algorithms, PSNR of each frame is plotted; $T=8$ is used.}}
	\label{fig:Traffic_cacti}
\end{figure}

\vspace{-2mm}
Next, we verify the proposed GAP-TV algorithm for the video compressive sensing with the regime followed by the coded aperture compressive temporal imaging (CACTI)~\cite{Patrick13OE}.
The GAP algorithm has been employed under this framework using wavelet and DCT transformation and has been demonstrated excellent performance for both grayscale and color video~\cite{Yuan14CVPR} compressive sensing.
By imposing the tree structure weights corresponding to the wavelet, GAP has demonstrated improved performance~\cite{Yuan14CVPR} compared with the original version proposed in~\cite{Liao14GAP}.
However, it is very tricky to select the groups and group  weights in the transform domain.
Therefore, in this paper, we utilize the GAP-TV algorithm, which is almost parameter-free, for the video CS as used in CACTI.
The sensing process of CACTI can be formulated as:
\begin{equation}
\textstyle{\Ymat^{(m)} = \sum_{f=1}^T \Phimat^{(f)} \odot \Xmat^{(m,f)}}
\end{equation} 
where $\Xmat^{(m,f)}\in {\mathbb R}^{L_x \times L_y}$ is the $f$-th frame corresponding to $m$-th measurement, $\Ymat^{(m)}\in {\mathbb R}^{L_x \times L_y}$; $\Phimat^{(f)}\in \{0,1\}^{L_x \times L_y}$ is the shifting binary mask used to encode $f$-th frame; each $\Phimat^{(f)}$ is a shifted version of others. $\odot$ denotes the Hadamard (element-wise) product.
Simulation results of GAP-TV compared with other algorithms are shown in Figure~\ref{fig:Traffic_cacti}.

\vspace{-4mm}
\subsection{Hyperspectral Image CS}
\label{Sec:hsi_cs}
\begin{figure}[htbp]
	\centering
	\vspace{-3mm}
	\includegraphics[width=.45\textwidth]{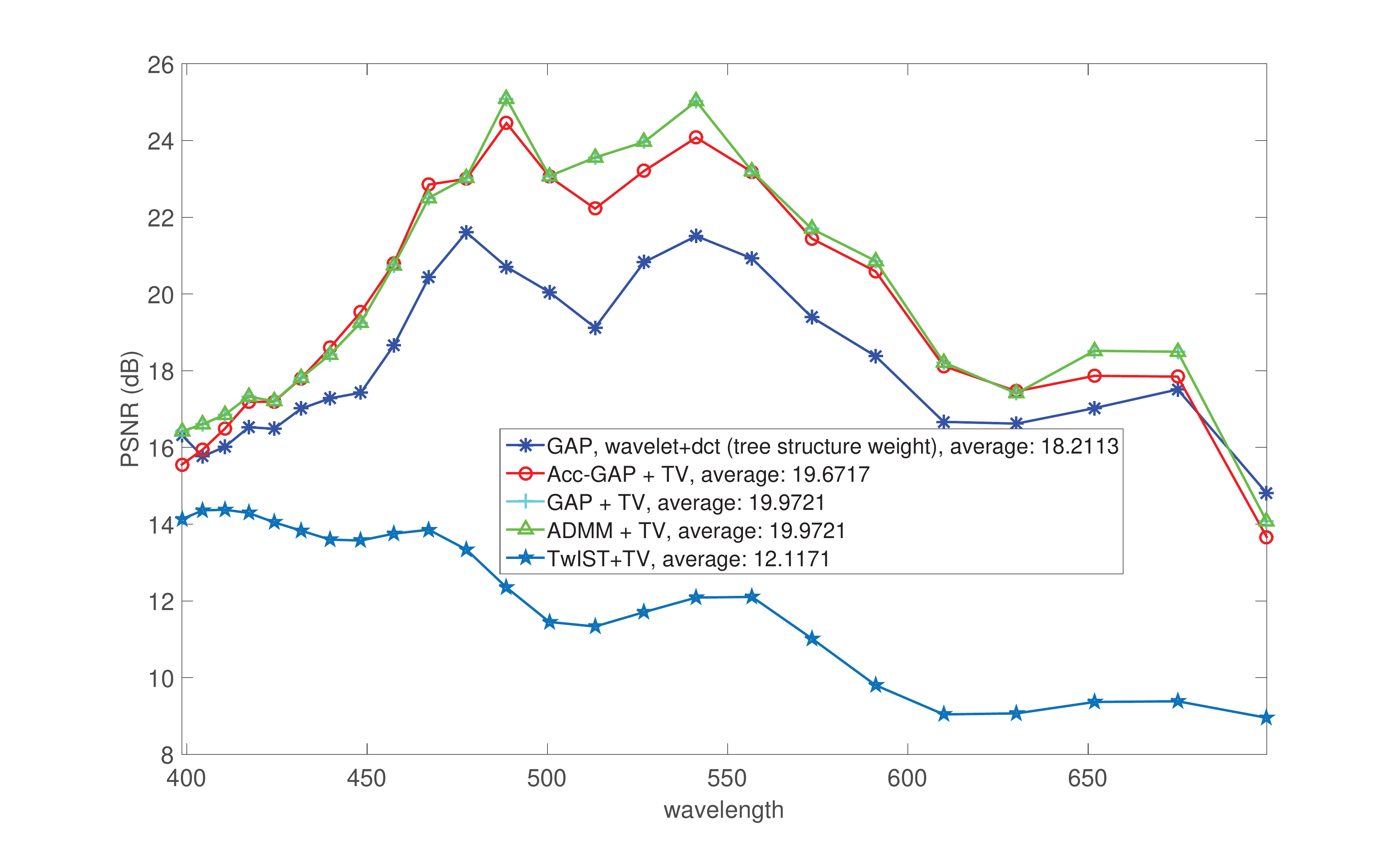}
	\vspace{-5mm}
	\caption{\small{Reconstruction PSNR of each frame for ``bird" with different algorithms.}}
	\label{fig:bird_psnr_sim}
\end{figure}

\vspace{-2mm}
For the hyperspectral image CS, we follow the architecture in the coded aperture snapshot spectral imaging (CASSI)~\cite{Wagadarikar09CASSI,Tsai15OL}, where the formulation is similar to CACTI discussed in Section~\ref{Sec:video_cs}. Each frame at different bandwidth is modulated by a shifting version of the physical mask.
We conduct the simulation on the ``bird" data as used in~\cite{Yuan15JSTSP} with results shown in Figure~\ref{fig:bird_psnr_sim}.

\vspace{-6mm}
\section{Conclusion}
\vspace{-2mm}
We have developed the generalized alternating projection based total variation minimization algorithm for compressive sensing. Excellent results have been demonstrated on compressive sensing of images, videos and hyperspectral images.  
%
%
\small
\bibliographystyle{IEEEbib}


\begin{thebibliography}{10}
	
	\bibitem{Liao14GAP}
	X.~Liao, H.~Li, and L.~Carin,
	\newblock ``Generalized alternating projection for weighted-$\ell_{2,1}$
	minimization with applications to model-based compressive sensing,''
	\newblock {\em SIAM Journal on Imaging Sciences}, vol. 7, no. 2, pp.
	797–--823, 2014.
	
	\bibitem{Yuan15Lensless}
	X.~Yuan, H.~Jiang, G.~Huang, and P.~Wilford,
	\newblock ``Lensless compressive imaging,''
	\newblock {\em arXiv:1508.03498}, 2015.
	
	\bibitem{Yuan14TSP}
	X.~Yuan, V.~Rao, S.~Han, and L.~Carin,
	\newblock ``Hierarchical infinite divisibility for multiscale shrinkage,''
	\newblock {\em IEEE Transactions on Signal Processing}, vol. 62, no. 17, pp.
	4363--4374, Sep. 1 2014.
	
	\bibitem{Yuan15GMM}
	X.~Yuan, H.~Jiang, G.~Huang, and P.~Wilford,
	\newblock ``Compressive sensing via low-rank {Gaussian} mixture models,''
	\newblock {\em arXiv:1508.06901}, 2015.
	
	\bibitem{Yuan15GAP}
	X.~Yuan, H.~Jiang, and P.~Wilford,
	\newblock ``Convergence of the generalized alternating projection algorithm for
	compressive sensing,''
	\newblock {\em arXiv:1509.06253}, 2015.
	
	\bibitem{Yuan15JSTSP}
	X.~Yuan, T.-H. Tsai, R.~Zhu, P.~Llull, D.~J. Brady, and L.~Carin,
	\newblock ``Compressive hyperspectral imaging with side information,''
	\newblock {\em IEEE Journal of Selected Topics in Signal Processing}, vol. 9,
	no. 6, pp. 964--976, September 2015.
	
	\bibitem{Tsai15OL}
	T.-H. Tsai, P.~Llull, X.~Yuan, D.~J. Brady, and L.~Carin,
	\newblock ``Spectral-temporal compressive imaging,''
	\newblock {\em Optics Letters}, vol. 40, no. 17, pp. 4054--4057, Sep 2015.
	
	\bibitem{Llull13COSI}
	P.~Llull, X.~Liao, X.~Yuan, J.~Yang, D.~Kittle, L.~Carin, G.~Sapiro, and
	D.~Brady,
	\newblock ``Compressive sensing for video using a passive coding element,''
	\newblock in {\em Imaging and Applied Optics}, 2013, p. CM1C.3.
	
	\bibitem{Patrick13OE}
	P.~Llull, X.~Liao, X.~Yuan, J.~Yang, D.~Kittle, L.~Carin, G.~Sapiro, and D.~J.
	Brady,
	\newblock ``Coded aperture compressive temporal imaging,''
	\newblock {\em Optics Express}, pp. 698--706, 2013.
	
	\bibitem{Yang14GMM}
	J.~Yang, X.~Yuan, X.~Liao, P.~Llull, G.~Sapiro, D.~J. Brady, and L.~Carin,
	\newblock ``Video compressive sensing using {G}aussian mixture models,''
	\newblock {\em IEEE Transaction on Image Processing}, vol. 23, no. 11, pp.
	4863--4878, November 2014.
	
	\bibitem{Yang14GMMonline}
	J.~Yang, X.~Liao, X.~Yuan, P.~Llull, D.~J. Brady, G.~Sapiro, and L.~Carin,
	\newblock ``Compressive sensing by learning a {G}aussian mixture model from
	measurements,''
	\newblock {\em IEEE Transaction on Image Processing}, vol. 24, no. 1, pp.
	106--119, January 2015.
	
	\bibitem{Tsai15COSI}
	T.H. Tsai, P.~Llull, X.~Yuan, L~Carin, and D.J. Brady,
	\newblock ``Coded aperture compressive spectral-temporal imaging,''
	\newblock in {\em Computational Optical Sensing and Imaging (COSI)}, 2015, pp.
	1--3.
	
	\bibitem{Yuan15FiO}
	X.~Yuan and S.~Pang,
	\newblock ``Structured illumination temporal compressive microscopy,''
	\newblock in {\em Frontier in Optics (FiO)}, 2015.
	
	\bibitem{Stevens15ASCI}
	A.~Stevens, L.~Kovarik, P.~Abellan, X.~Yuan, L.~Carin, and N.~D. Browning,
	\newblock ``Applying compressive sensing to tem video: A substantial framerate
	increase on any camera,''
	\newblock {\em Advanced Structural and Chemical Imaging}, 2015.
	
	\bibitem{CS15Book}
	P.~Llull, X.~Yuan, X.~Liao, J.~Yang, D.~Kittle, L.~Carin, G.~Sapiro, and D.J.
	Brady,
	\newblock ``Temporal compressive sensing for video,''
	\newblock in {\em Compressed Sensing and its Applications}, H.~Boche,
	R.~Calderbank, G.~Kutyniok, and J.~Vybíral, Eds., pp. 41--74. Springer
	International Publishing, 2015.
	
	\bibitem{Yuan14CVPR}
	X.~Yuan, P.~Llull, X.~Liao, J.~Yang, G.~Sapiro, D.~J. Brady, and L.~Carin,
	\newblock ``Low-cost compressive sensing for color video and depth,''
	\newblock in {\em IEEE Conference on Computer Vision and Pattern Recognition
		(CVPR)}, 2014.
	
	\bibitem{Llull14COSI}
	P.~Llull, X.~Yuan, X.~Liao, J.~Yang, L.~Carin, G.~Sapiro, and D~Brady,
	\newblock ``Compressive extended depth of field using image space coding,''
	\newblock in {\em Computational Optical Sensing and Imaging (COSI)}, 2014, pp.
	1--3.
	
	\bibitem{Llull15Optica}
	P.~Llull, X.~Yuan, L.~Carin, and D.~Brady,
	\newblock ``Image translation for single-shot focal tomography,''
	\newblock {\em Optica}, 2015.
	
	\bibitem{Tsai15OE}
	T.-H. Tsai, X.~Yuan, and D.~J. Brady,
	\newblock ``Spatial light modulator based color polarization imaging,''
	\newblock {\em Optics Express}, vol. 23, no. 9, pp. 11912--11926, May 2015.
	
	\bibitem{Bioucas-Dias2007TwIST}
	J.M. Bioucas-Dias and M.A.T. Figueiredo,
	\newblock ``A new {TwIST}: Two-step iterative shrinkage/thresholding algorithms
	for image restoration,''
	\newblock {\em IEEE Transactions on Image Processing}, vol. 16, no. 12, pp.
	2992--3004, December 2007.
	
	\bibitem{Wang08TV}
	Y.~Wang, J.~Yang, W.~Yin, and Y.~Zhang,
	\newblock ``A new alternating minimization algorithm for total variation image
	reconstruction,''
	\newblock {\em SIAM Journal on Imaging Sciences}, vol. 1, no. 3, pp. 248--272,
	2008.
	
	\bibitem{Yang10TV}
	J.~Yang, Y.~Zhang, and W.~Yin,
	\newblock ``A fast alternating direction method for tvl1-l2 signal
	reconstruction from partial fourier data,''
	\newblock {\em IEEE Journal of Selected Topics in Signal Processing}, vol. 4,
	no. 2, pp. 288--297, 2010.
	
	\bibitem{Li13COA}
	C.~Li, W.~Yin, H.~Jiang, and Y.~Zhang,
	\newblock ``An efficient augmented lagrangian method with applications to total
	variation minimization,''
	\newblock {\em Computational Optimization and Applications}, vol. 56, no. 3,
	pp. 507--530, 2013.
	
	\bibitem{Huang13ICIP}
	G.~Huang, H.~Jiang, K.~Matthews, and P.~Wilford,
	\newblock ``Lensless imaging by compressive sensing,''
	\newblock {\em IEEE International Conference on Image Processing}, 2013.
	
	\bibitem{Jiang14APSIPA}
	H.~Jiang, G.~Huang, and P.~Wilford,
	\newblock ``Multi-view in lensless compressive imaging,''
	\newblock {\em APSIPA Transactions on Signal and Information Processing}, vol.
	3, no. 15, pp. 1--10, 2014.
	
	\bibitem{Yuan14Tree}
	X.~Yuan, P.~Llull, D.~Brady, and L.~Carin,
	\newblock ``Tree-structure bayesian compressive sensing for video,''
	\newblock {\em arXiv:1410.3080}, 2014.
	
	\bibitem{Beck09TV}
	A.~Beck and M.~Teboulle,
	\newblock ``Fast gradient-based algorithms for constrained total variation
	image denoising and deblurring problems,''
	\newblock {\em IEEE Transactions on Image Processing}, vol. 18, no. 11, pp.
	2419--2434, Nov. 2009.
	
	\bibitem{zhu08Dual}
	M.~Zhu, S.~J. Wright, and T.~F. Chan,
	\newblock ``{Duality-based algorithms for total-variation-regularized image
		restoration},''
	\newblock {\em Computational Optimization and Applications}, 2008.
	
	\bibitem{ADMM2011Boyd}
	Stephen Boyd, Neal Parikh, Eric Chu, Borja Peleato, and Jonathan Eckstein,
	\newblock ``Distributed optimization and statistical learning via the
	alternating direction method of multipliers,''
	\newblock {\em Found. Trends Mach. Learn.}, vol. 3, no. 1, pp. 1--122, January
	2011.
	
	\bibitem{Dong14TIP}
	W.~Dong, G.~Shi, X.~Li, Y.~Ma, and F.~Huang,
	\newblock ``Compressive sensing via nonlocal low-rank regularization,''
	\newblock {\em IEEE Transactions on Image Processing}, vol. 23, no. 8, pp.
	3618--3632, 2014.
	
	\bibitem{Jiang12Inverse}
	H.~Jiang, W.~Deng, and Z.~Shen,
	\newblock ``Surveillance video processing using compressive sensing,''
	\newblock {\em Inverse Problems and Imaging}, vol. 5, no. 2, pp. 201--214,
	2012.
	
	\bibitem{Wagadarikar09CASSI}
	A.~Wagadarikar, N.~Pitsianis, X.~Sun, and David Brady,
	\newblock ``Video rate spectral imaging using a coded aperture snapshot
	spectral imager,''
	\newblock {\em Optics Express}, vol. 17, no. 8, pp. 6368--6388, 2009.
	
\end{thebibliography}

\end{document}